\documentclass[mathleft,fleqn,%
usenames,dvipsnames,%
]{an-arxiv}
\usepackage{graphicx}
\usepackage[varg]{txfonts}
\usepackage{natbib}
\bibpunct{(}{)}{;}{a}{}{,}
\usepackage{amssymb}
\usepackage{comment}

\makeatletter
\newcommand{\ionsc}[2]{#1\,\textsc{\rmfamily\@roman{#2}}\relax}
\makeatother

\newcommand{\msigma}{$\mbh$--$\sigma$}

\newcommand{\mbh}{M_\mathrm{BH}}
\newcommand{\sigmab}{\sigma_\mathrm{b}}
\newcommand{\Pjet}{P_\mathrm{jet}}
\newcommand{\nISM}{n_\mathrm{ISM}}
\newcommand{\vcr}{\langle v_\mathrm{c,r}\rangle}
\newcommand{\Rcmax}{R_\mathrm{c,max}}
\newcommand{\Rbulge}{R_\mathrm{bulge}}
\newcommand{\taujc}{\tau_\mathrm{jc}}
\newcommand{\fvol}{f_\mathrm{V}}
\newcommand{\eddcrit}{\lambda_\mathrm{crit}}
\newcommand{\sfrsd}{Sectionigma_\mathrm{SFR}}

\newcommand{\kpc}{\,\mathrm{kpc}}

\newcommand{\kms}{\,\mathrm{km}\,\mathrm{s}^{-1}}
\newcommand{\cmq}{\,\mathrm{cm}^{-3}}
\newcommand{\ergs}{\,\mathrm{erg}\,\mathrm{s}^{-1}}
\newcommand{\sr}{\,\mathrm{sr}}

\begin{document}

\Pagespan{1}{}
\Yearpublication{2014}%
\Yearsubmission{2014}%
\Month{0}%
\Volume{999}%
\Issue{0}%
\DOI{asna.201400000}%

\title{Galaxy-scale AGN Feedback - Theory}

\author{A.\,Y. Wagner\inst{1}\fnmsep\thanks{Corresponding author:
  \email{ayw@ccs.tsukuba.ac.jp}}
\and  G.\,V. Bicknell\inst{2}
\and  M. Umemura\inst{1}
\and  R.\,S. Sutherland\inst{2}
\and J.\,Silk\inst{3,4,5,6}
}
\titlerunning{Feedback - Theory}
\authorrunning{A.\,.Y. Wagner et al.}
\institute{
University of Tsukuba, Center for Computational Sciences, Tennodai 1-1-1, Tsukuba, Ibaraki, Japan
\and 
The Australian National University, Research School of Astronomy \&{} Astrophysics, Weston Creek, ACT 2611, Australia
\and
Institut d'Astrophysique de Paris (UMR 7095: CNRS \&{} UPMC -- Sorbonne Universit\'{e}s), 98 bis bd Arago, F-75014 Paris, France
\and
Laboratoire AIM-Paris-Saclay, CEA/DSM/IRFU, Univ. Paris VII, F-91191 Gif-sur-Yvette, France
\and
Department of Physics and Astronomy, The Johns Hopkins University Homewood Campus, Baltimore, MD 21218, USA
\and
BIPAC, Department of Physics, University of Oxford, Keble Road, Oxford OX1 3RH, UK
}



\abstract{%
Powerful relativistic jets in radio galaxies are capable of driving strong outflows but also inducing star-formation by pressure-triggering collapse of dense clouds. We review theoretical work on negative and positive active galactic nuclei feedback, discussing insights gained from recent hydrodynamical simulations of jet-driven feedback on galaxy scales that are applicable to compact radio sources. The simulations show that the efficiency of feedback and the relative importance of negative and positive feedback depends strongly on interstellar medium properties, especially the column depth and spatial distribution of clouds. Negative feedback is most effective if clouds are distributed spherically and individual clouds have small column depths, while positive feedback is most effective if clouds are predominantly in a disc-like configuration.}

\maketitle

\section{Introduction - The challenges of constructing theories of AGN feedback}

The energy released by Active Galactic Nuclei (AGN) into their host galaxies' interstellar medium (ISM) in the form of relativistic jets, fast winds, and radiation, is believed to have a strong impact on the host galaxies' evolution \citep{Alexander2012a, Fabian2012a, Taylor2015a}. However, a satisfactory theory of the inner workings and the efficiency of AGN feedback does not yet exist.

The interactions between AGN outflows and the ISM were at first investigated mainly in order to explain the emission lines and kinematics of broad and narrow line region clouds in quasars \citep{Blumenthal1979a, Weymann1982a, Schiano1986a}. Most of the work focused on ``quasar winds'' presumed to be driven by radiation pressure \citep{Williams1972a, Mushotzky1972a}. The full realization of the importance of AGN outflows in the coevolution of supermassive black holes (SMBH) and their host galaxies came through the theory of the quasar wind establishing the \msigma{} relation \citep{Silk1998a}. The role of radio jets in galaxy evolution at that point was unclear, although cases in which jets drive outflows \citep{Morganti1998a} and cases in which jets might be responsible for induced star-formation were reported \citep{Begelman1989a, vanBreugel1993a, Bicknell2000a}, starting with the alignment effect \citep{McCarthy1987a, Chambers1987a}. Interactions of radio jets with the ISM were often alluded to in the context of bent jet morphologies \citep{Eilek1984a, Higgins1995a}, but also in the context of the properties of emission line regions \citep{Heckman1981a, Capetti1996a, Steffen1997a}. The strongest link between radio jets and galaxy evolution came from the requirement to overcome catastrophic cooling in galaxy cluster halos \citep{Mathews2003a}, and keep the massive, early-type galaxies passively star-forming \citep{Benson2003a}. The idea that radio jets may affect galaxy evolution in a similar manner to quasar winds is a more recent idea, motivated not least by the strong interactions between jet and ISM on kpc scales that appear be occurring in compact radio sources \citep{Morganti2005a, Holt2008a}.

The lack of understanding of the inner workings and the efficiency of AGN feedback comes primarily from two complications in the numerical modelling of AGN feedback. One is the inability to capture the wide range of physical scales involved in the AGN feedback cycle. It is likely that cycles operate on multiple scales, which makes it difficult to capture both inflows, that is, accretion toward the black hole, and outflows (jets, winds) within the same model. Fluid-dynamical simulations need to be conducted in three dimensions in order to capture filamentary accretion and clumpy, fragmented outflows realistically \citep{Gaspari2011b}. However, capturing a dynamic range of more than 10 orders of magnitude from the Schwarzschild radius to hundreds of kpc from the central supermassive black hole (SMBH) is not feasible in three-dimensional simulations, even with adaptive mesh refinement or smoothed-particle hydrodynamics codes. Independent simulations on different scales need to be performed and matched at their boundaries, e.g., as in zoom simulations \citep{Dubois2015a}. 

The other complication is the inadequate treatment of the ISM in models of AGN feedback. The ISM must be treated as a multiphase inhomogeneous medium, which again requires high-resolution three-dimensional simulations. The two complications cited above, the wide range of physical scales and an inhomogeneous ISM, are the reason why a pc- to kpc-scale theory of AGN feedback has not been incorporated into cosmological simulations. Cosmological simulations typically have maximum spatial resolutions of $1\kpc$, a density range of only up to $\sim100\cmq$, and multiphase gas and sub-kpc physics can only be included through sub-grid models \citep{Barai2013a}.

A third problem in understanding AGN feedback is that we do not know exactly what happens at the apex of the feedback cycle near the SMBH: What accretion flow and SMBH properties lead to what kind of outflows? This is addressed in the contribution by Czerny et al. in this Volume, and is also related to the unknown duty cycle of AGN jets.

In the following, we first touch on the topic of SMBH--bulge coevolution in Section~\ref{s:bhbc}, since this is one of the main motivations to study AGN feedback, and then list the various types (or modes) of feedback frequently mentioned in the study of galaxy formation in Section~\ref{s:tof}. In Section~\ref{s:sims} we explain, through hydrodynamical simulations, how negative and positive jet-driven AGN feedback works, highlighting the dependence of the feedback efficiency on ISM properties. In Section~\ref{s:summary} we conclude with a brief summary.

\section{Black-hole bulge coevolution}\label{s:bhbc}

The tight correlations between the SMBH mass, $\mbh$ and properties of the classical bulge of the host galaxy, including its mass, velocity dispersion $\sigmab$, and luminosity \citep{kormendy2013a}, is at the core of the concept of coevolution of SMBHs and galaxy bulges. There are three possibilities to explain these correlations: i) the SMBH controls the growth of the bulge through negative feedback \citep[e.g.,][]{Silk1998a} or through positive feedback \citep{Ishibashi2012a,Silk2013a}; ii) the bulge controls the growth of the SMBH \citep{Umemura2001a}; iii) SMBH and bulge masses are averaged statistically through successive galaxy mergers \citep[e.g., ][]{Jahnke2011a}. These three possibilities are not mutually exclusive, so that, given the plenitude of processes predicting $\mbh$--bulge correlations, it is perhaps not too surprising that such correlations should come into existence. However, the relative importance of these possibilities is not known because the efficiencies with which they operate have not yet been thoroughly investigated.

In this contribution, we are primarily concerned with the first of the above possibilities, viz. the SMBH controls the growth of the bulge via outflows generated near the environs of the SMBH. The fundamental ideas of the theory in the case of negative feedback were first formulated by \citet{Silk1998a} through a one-dimensional analytic description of a quasar wind, modelled as an energy-driven, spherical outflow propagating through an initially hydrostatic isothermal bulge. The condition imposed for sufficient feedback was that the speed of the shell of swept-up material exceed the velocity dispersion of the bulge. This condition led to a linear relation between $\mbh$ and $\sigmab^5$. As was recognised for stellar winds by \citet{Dyson1984a}, AGN-powered outflows can be energy-driven or momentum-driven. A condition for sufficiently strong feedback in the case of momentum-driven outflows was derived by \citet{Fabian1999a} and \citet{King2003a}, yielding a linear relation between $\mbh$ and $\sigmab^4$ with no free parameters and approximately correct normalisation.

\section{Types of feedback}\label{s:tof}

Many classifications of feedback are in use that are often contrasted in pairs \citep[see also][]{Combes2015a}: ``Quasar-mode'' versus ``Radio-mode'' feedback; ``Energy-driven'' versus ``Momentum-driven'' outflows; ``Mechanical'' or ``Kinetic'' versus ``Radiative'' feedback; ``Positive'' versus ``Negative'' feedback; ``'Maintenance-mode'' versus ``Establishment-mode'' feedback. To go into details of all definitions is beyond the scope of this paper, and at any rate, there is no complete consensus on the definitions of the terminology, so we merely explain briefly the types and modes of feedback most relevant to compact radio sources: The interaction of young jets in a radio galaxy with the ISM leads to quasar-mode, energy-driven, mechanical, negative or positive feedback. 

AGN feedback by relativistic jets occurs from kpc galaxy scales \citep{Wagner2011a}, 
on which the jet interacts with the ISM, to Mpc scales \citep{Perucho2014a, Nawaz2014a}, on which the jet interacts with the intergalactic medium or the intra-cluster medium (ICM). The evolution and deceleration of a jet across these scales has been studied by, e.g., \citet{Cielo2014a} and \citet[][see also the contribution by Perucho et al. in this Volume]{Perucho2014b}. 
Feedback due to well-developed jets of massive radio galaxies that gently heat the ICM, suppress cooling flows, and thereby prevent the excessive growth of massive galaxies in line with the steepening of the galaxy luminosity function at the high mass end, is termed radio-mode feedback \citep{Mcnamara2012a}, and now also commonly maintenance mode feedback. Quasar-mode feedback is often used to describe galaxy-scale outflows observed in high-redshift quasars, possibly driven by radiation or disc winds. Feedback due to a young, confined jet dispersing, heating, or blowing out gas on scales $\sim1\kpc$ typical of the extent of Compact Steep Spectrum (CSS) or Gigahertz Peaked Spectrum (GPS) objects \citep{Mahony2013a, Dasyra2015a} is also a form of quasar-mode feedback since it occurs on galaxy scales and does not necessarily result in the heating of the host galaxy's extended halo or the ICM.


Feedback by relativistic jets is mediated by the jets' ram pressure and thermal pressure rather than purely by radiation and is therefore sometimes termed mechanical or kinetic feedback, as opposed to radiative feedback. The outflow induced by AGN jet feedback is in the energy-driven regime on all scales, meaning that radiative losses are small enough not to influence the dynamics of the outflow and its expanding, confining bubble \citep{Wagner2012a}.

Evidence for energy conservation in quasar-mode feedback is growing. Observational evidence for AGN accretion disc winds at $\sim0.2\,c$, which drive large-scale massive molecular outflows, have recently been detected in two low redshift ultraluminous infrared galaxies \citep{Tombesi2015b, Feruglio2015a}. The disk winds are launched from sub-mpc scales as evidenced by the time-variable, broad blue-shifted \ionsc{Fe}{25} absorption features. Momentum-driven flows fail to drive the observed fast ($\sim1000\kms$) molecular outflows detected on kpc scales, but energy-conserving outflows are found to provide the observed mechanical energy \citep{Faucher-Giguere2012a, Wagner2013a}.

There is reason to believe the feedback by AGN jets may be both positive, that is, it enhances star formation, and negative, that is, it inhibits star formation  during different or overlapping phases. On the one hand, the recent star formation and the alignment of emission line regions in many young radio galaxies may hint at positive feedback \citep{Privon2008a, Tadhunter2011a}. On the other hand, any gas dispersed, heated, or blown away through the interaction of the jet with the radio galaxy is likely to suppress star formation \citep{Morganti2005a, Nesvadba2008a}.

Finally, the feedback in CSS/GPS sources may, especially in gas-rich galaxies at high redshift, contribute to the establishment of $\mbh$--bulge relations (``establishment-mode'' feedback), or may, e.g. in recurring sources (\citealt{Shulevski2015a}; see also contribution by Brienza et al. in this Volume) or comparatively low-density, low-redshift sources aid, in maintaining $\mbh$--bulge relations (another variation of ``maintenance-mode'' feedback).

\section{Numerical simulations of galaxy-scale feedback}\label{s:sims}

\subsection{Jet-driven outflows}\label{s:jet-outflows}

\begin{figure*}
   \centering
   \includegraphics[width=\textwidth]{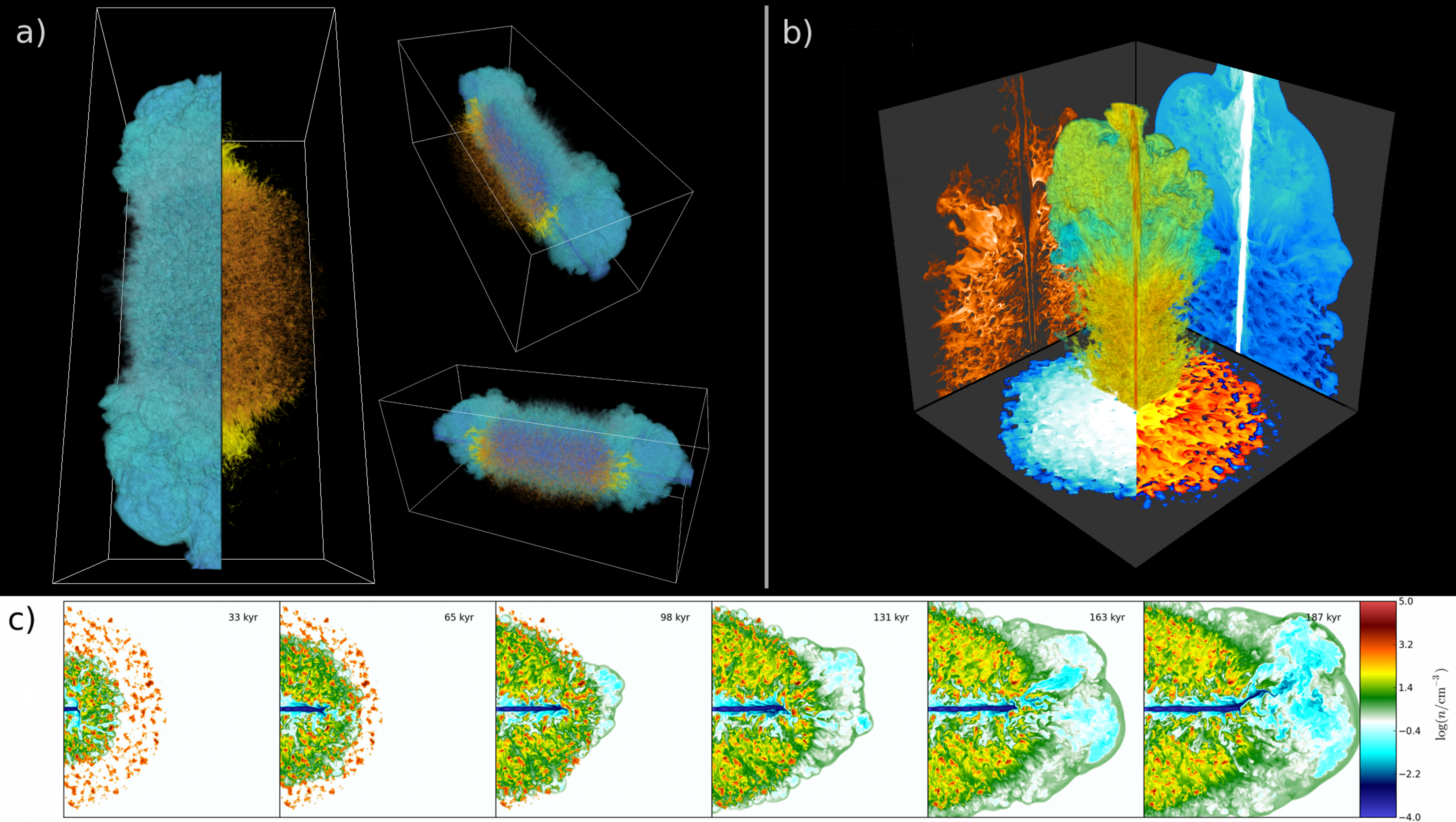} 
   \caption{Three-dimensional relativistic hydrodynamic simulations of AGN jets interacting with the ISM of a forming galaxy. a) Volume render at various angles of the jet in blue shown in one half of the galaxy, and the dispersed clouds in orange (low velocities of $\sim100$ to $\sim1000\kms$) and yellow (high velocities of $\gtrsim1000\kms$) shown in the other half; b) Volume render of the jet plasma (central column), and projections of mid-plane slices of various physical quantities to the box faces. Back-left: kinetic energy of clouds, back-right: velocity map, bottom-left: thermal pressure map, bottom-right: temperature map; c) Mid-plane slices of the density showing the evolution of the jet through the two-phase ISM.}
   \label{f:jet-ism}
\end{figure*}

The work by \citet{Silk1998a} inspires the kind of simulations that could be performed to study the physics of quasar-mode AGN feedback on galaxy scales. Motivated by the possibility of jet-driven feedback in young, gas-rich radio-galaxies, we conducted a series of three-dimensional relativistic hydrodynamic simulations investigating the dependence of AGN feedback on the jet power and the properties of the ISM \citep{Wagner2012a}. The ISM was initialised as a two-phase fractal gas distribution consisting of a warm, dense phase of clouds with a given filling factor and maximum size, whose density ranged from a few 10s cm$^{-3}$ to a few $10^5$ cm$^{-3}$ and obeyed a log-normal distribution, embedded in a hot, diffuse galactic halo. The jet parameters were chosen to be in the range $10^{43}\ergs$ to $10^{46}\ergs$, typical of powerful AGN jets, and the ISM parameters were typical of clumpy, gas-rich high-redshift forming galaxies \citep{Shapley2011a, Carilli2013a}. The focus was on the efficiency of energy transfer from the jet to the dense clouds, since it is there that the bulk of the star formation occurs.

Figure~\ref{f:jet-ism} shows various visualisations of the simulations. The first realisation from these simulations was that the jet, while percolating through the channels of the porous clouds, dispersed clouds substantially at all radii over $4\pi\sr$ solid angle, indicating that AGN jet feedback does have an effect on the dense phase in the ISM, e.g., atomic and molecular clouds. As the jet struggled through the dense field of clouds, it was deflected, split, and confined, because the jet density was typically eight orders of magnitude lower than the density of the clouds. In the process, the jet blew an energy-driven bubble through the ISM. The forward shock swept up the hot ISM and the surface of the clouds, disrupting them slightly. The primary mode of energy transfer occurred through the channel flows, which were mass-loaded through the hot ISM and gradually through surface ablation of the clouds, and carried substantial ram pressure (comparable to or a factor of a few greater than the thermal pressure). The hydrodynamic ablation of clouds was driven by the shear with the surrounding flow, which generates Kelvin-Helmholtz instabilities at the cloud interfaces. Clouds were also accelerated in bulk through direct impact of the channel flows.

The above describes the generic inner workings of negative, galaxy-scale, mechanical feedback. It is the same for AGN jets and disc winds, as long as energy is largely conserved \citep{Wagner2013a}. We found that the strength of this form of feedback, which was quantified as the density-averaged radial outflow velocity of the clouds, $\vcr$, after their interaction with the jet, was very high. As in \citet{Silk1998a}, if the outflow velocity $\vcr$ was larger than the value of $\sigmab$ predicted by the \msigma{} relation, then feedback was deemed efficient. The critical Eddington ratio of a jet, $\eddcrit$, below which it cannot disperse the dense clouds of a galaxy to $\sigmab$, depends strongly on ISM parameters. For example, for a $10^{45}\ergs$ jet, if cloud sizes were $\sim10$ pc, $\eddcrit\sim10^{-4}$, if cloud sizes were $\sim25$ pc, $\eddcrit\sim10^{-3}$, and if cloud sizes were $\sim50$ pc, $\eddcrit\sim10^{-1}$.

\begin{figure}[]
   \centering
   \includegraphics[width=\linewidth]{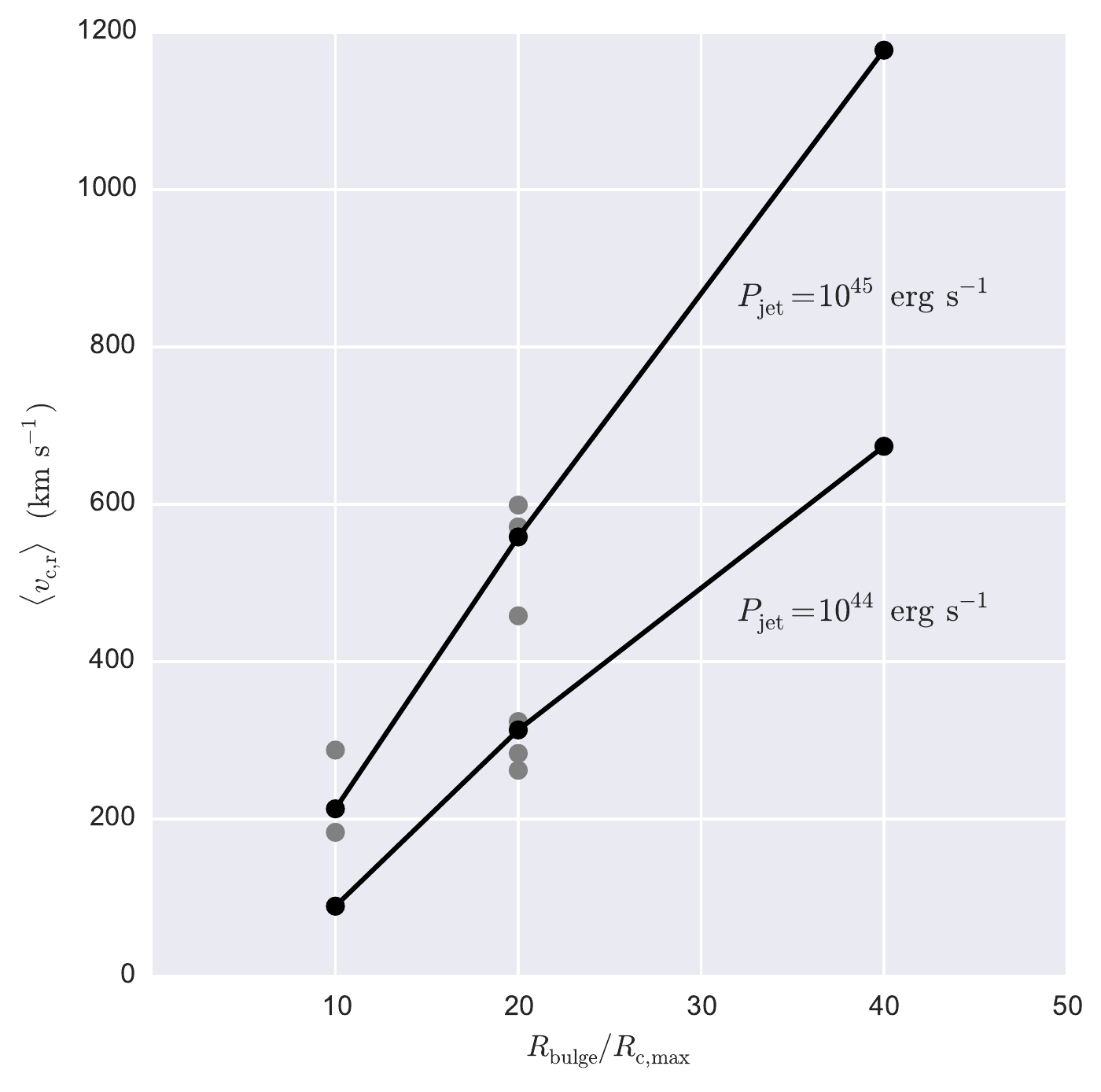} 
   \caption{The almost linear dependence of the strength of negative feedback, $\vcr$, on the inverse of the maximum cloud size, $\Rbulge/\Rcmax$, for a fixed bulge radius, $\Rbulge$. Each point is the result from one simulation. The black points are simulations with fixed volume filling factor and ISM density and points connected are simulations with the same jet power. The grey points are other simulations with the same parameters except for the volume filling factor.}
   \label{f:vcr_vs_rmax}
\end{figure}

The most important parameter determining the efficiency of negative feedback was the maximum size of clouds, or, equivalently, their maximum column density. For a fixed volume filling factor, $\fvol$, and total mass, the smaller the clouds, the easier they were to disperse because the ratio of surface area subjected to the KH instability and ablation increases in relation to the mass of the cloud. Conversely, bigger clouds were harder to disperse, but were, instead, more susceptible to collapse and induced star formation (see Section~\ref{s:negpos}). 

The dependence of the strength of negative feedback on cloud sizes can be extended to the scale of the bulge of a galaxy by approximating the propagation of the jet plasma through the clumpy ISM as diffusion of jet streams scattering against clouds. We defined the jet-cloud \lq\lq{}interaction depth\rq\rq{}, $\taujc$, for a given distribution of clouds, analogous to the optical depth of photons: $\taujc = n_c \pi \Rcmax^2 \Rbulge$, where $n_c$ is the number density of clouds, and $\Rbulge$ is the radius of the region in the bulge which contains clouds \citep{Wagner2012a} . The clouds may be thought of as $N$ scattering centers with a cross-section $\pi\Rcmax^2$. The total number of clouds in the bulge is $N=\fvol\Rbulge^3/\Rcmax^3=n_c\Rbulge^3$. Therefore, the number density of clouds is $n_c=\fvol/\Rcmax^3$, and the interaction depth is $\taujc=\pi\fvol(\Rbulge/\Rcmax)$. Hence, for fixed $\fvol$ and $\Rbulge$, $\taujc\propto\Rcmax^{-1}$. If $\vcr\propto\taujc$, then this explains the almost linear relationship seen in Figure~\ref{f:vcr_vs_rmax} between $\Rbulge/\Rcmax$ and $\vcr$.

Since the outflow remains energy-driven, the strength of negative feedback scaled with $\Pjet/\nISM$, and the mechanical advantage measured with respect to the dense phase, was $\gg 1$ despite the porosity of the dense phase.

While we have mainly focused on the ratio of outflow speed to velocity dispersion of the dense phase in the bulge as an indicator of the negative feedback efficiency, one could also look at the ratio of outflow speed to escape speed from the bulge. That is, how much of the dense phase can be removed from the galaxy potential? The escape speed from the bulge is typically a few times the velocity dispersion, so we may expect the dependence of the conditions for efficient feedback on the Eddington ratio of the jet and on the ISM parameters to be similar. However, this needs to be tested by following the fate of the outflowing material with simulations of feedback over much longer timescales (tens of Myrs) and on much larger spatial scales (tens of kpc) with the same resolution (a few pc) using adaptive mesh techniques. \citet{Bland-Hawthorn2011a} found that gas removal even from dwarf galaxies is very difficult, when considering a clumpy ISM, so we may encounter similar difficulties in removing large amounts of inhomogeneous gas from massive galaxies.

\subsection{Positive feedback} \label{s:negpos}

When might star formation in clouds be induced by pressure triggered collapse rather than inhibited by ablation, dispersal, and outflows? In Section~\ref{s:jet-outflows} we mentioned how the typical column density of the clouds affects the efficiency of negative feedback. Smaller clouds are more susceptible to hydrodynamic ablation and subsequent dispersion than bigger clouds, which are instead more susceptible to collapse and induced star formation. The jet-blown bubble (see Figure~\ref{f:jet-ism}) engulfs the clouds in a plasma with overpressure of a factor of up to 1000. This would reduce the critical Bonnor-Ebert mass of a cloud by a factor $\sim20$. A further constraint comes from the comparison between the ablation and collapse timescales of clouds, and the ablation timescale, although difficult to estimate, is proportional to the cloud radius. These considerations reinforce the relevance of cloud properties, especially the column density of clouds, in the assessment of whether feedback is positive or negative and how efficient either mode might be.

\begin{figure}[]
   \centering
   \includegraphics[width=\linewidth]{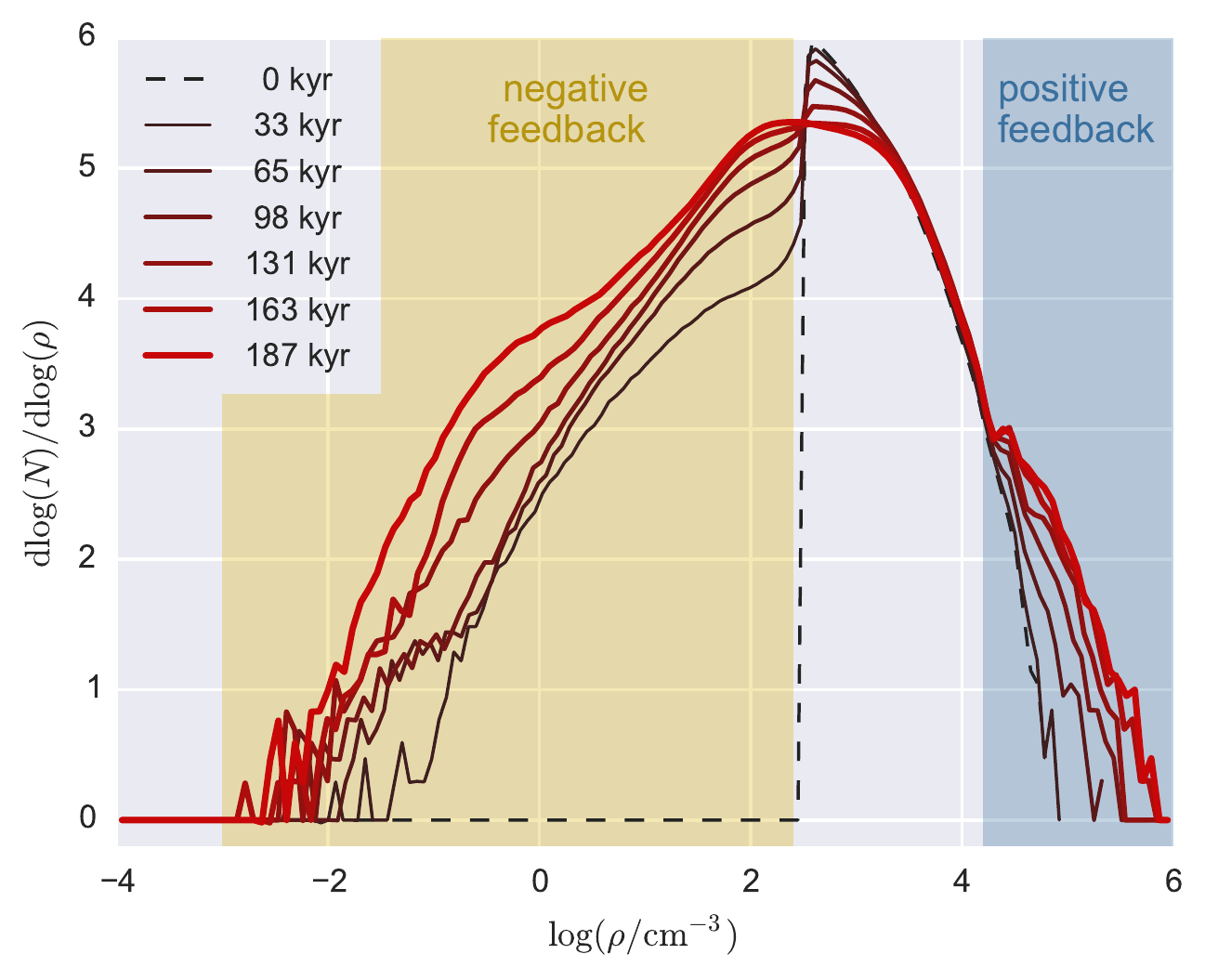} 
   \caption{The time-evolution of the density PDF of clouds in one of the simulations by \citet{Wagner2012a}. At $t>0$, both a region of diffuse gas ($\rho\lesssim$ few $10^2\cmq$) and a region of dense gas ($\rho\gtrsim$ few $10^4\cmq$) appear: the former is associated with negative feedback due to cloud ablation, and the latter is associated with positive feedback due to cloud compression.}
   \label{f:rhopdf}
\end{figure}

Negative and positive feedback may both occur simultaneously in the same galaxy during one phase of AGN activity. \citet{Cresci2015a} have found a possible example of mixed feedback, which observationally manifests itself as a high-dispersion, blue-shifted outflow cone surrounded by knots of star-formation bright in H$\alpha$ and in the UV continuum. In such cases, it is the relative efficiency of negative and positive feedback and the timescale on which they operate, which determine the net effect on that galaxy's evolution.

The simplest estimate of how much negative and how much positive feedback occurs is to measure the amount of cloud ablation and the amount of cloud coagulation over the time the AGN is active. In our simulations, we can see the effect of both through the time-evolution of the density probability distribution function (PDF) of the clouds, which is shown in Figure~\ref{f:rhopdf}. Material near the surface of clouds associated with the region of the peak of the log-normal density PDF\footnote{Note, the PDF initially has a lower density cutoff below which clouds are thermally unstable.} is ablated and forms a low-density wing of diffuse cloud material in the PDF. The integrated mass within this wing can be considered a measure of negative feedback. A high density wing also appears due to radiative shocks propagating into the clouds and cloud compression by the ambient pressure. The integrated mass under the high-density wing may be considered a measure of positive feedback.

\begin{figure}[]
   \centering
   \includegraphics[width=\linewidth]{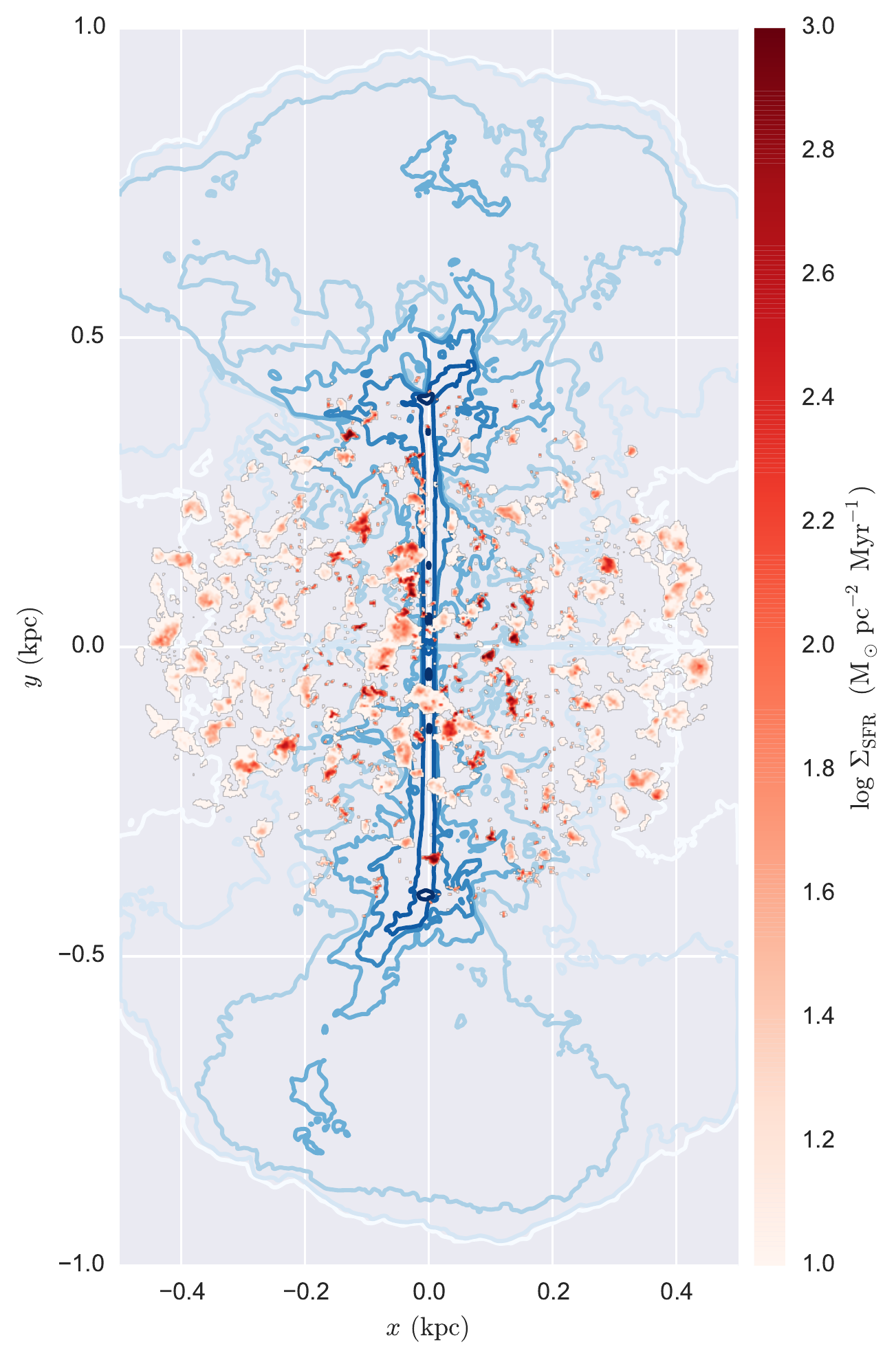} 
   \caption{Map of the SFR surface density, $\sfrsd$, overlaid on a synthetic radio surface brightness image of the jet plasma. The logarithmic contours of the radio surface brightness span six orders of magnitude and are arbitrarily normalised. The cores of dense, preferentially larger cloud complexes show enhanced star-formation aligned with the main jet stream. Occasionally, enhanced star-formation is seen at larger radii, e.g., at $(x, y) \approx (-0.2\kpc, -0.2\kpc)$, as a result of a secondary deflected jet stream impacting clouds.}
   \label{f:sfr}
\end{figure}

\begin{figure}[]
   \centering
   \includegraphics[width=\linewidth]{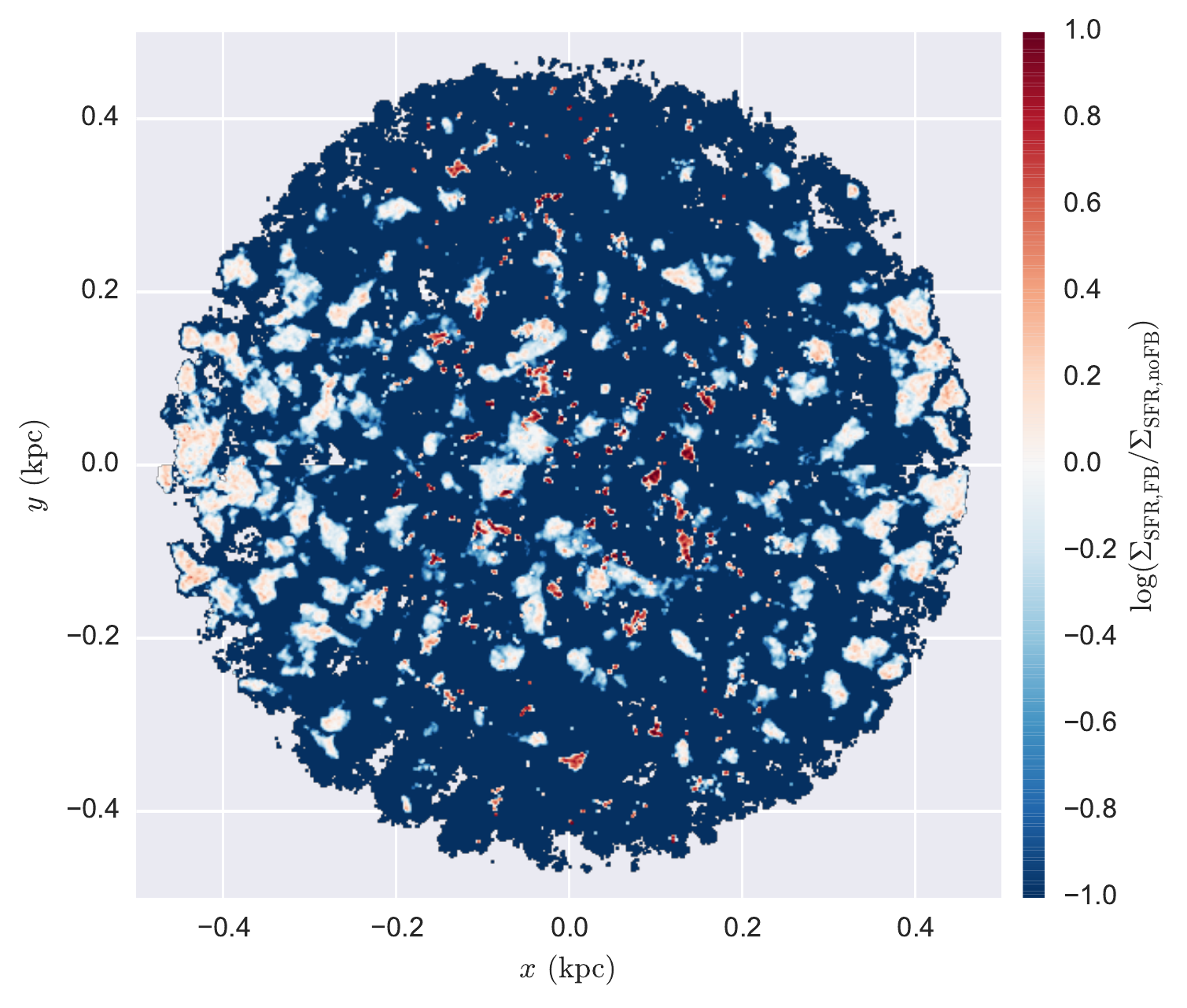} 
   \caption{Map of the ratio of SFR surface densities between the last and the first time steps of the simulation, that is, after feedback and before feedback. Red clumps are sites of jet-triggered star-formation, and in blue are  of the initial cloud distribution that were ablated away and experienced negative feedback. Jet-induced star-formation occurs preferentially in a cylindrical region about the jet axis, although some large clouds in the outer regions are compressed by the high-pressure cocoon and exhibit enhanced star-formation.}
   \label{f:sfrd}
\end{figure}

A more complicated evaluation of the relative importance of negative and positive feedback is to estimate the star-formation rate (SFR) as a function of time during an episode of AGN feedback. The main uncertainty here is that star-formation criteria are only approximate. A commonly employed formula is that the SFR in a region under consideration is proportional to the local density divided by the free-fall time. Additional criteria are sometimes applied, e.g., temperature and density thresholds and the requirement that the divergence of the flow velocity be negative. A map of the SFR surface density together with radio surface brightness contours from one of our simulations is shown in Figure~\ref{f:sfr}. An enhancement of the SFR due to compression by the jet can be seen in the cores of clouds aligned with the jet-cloud axis. Secondary jet streams that impact clouds at larger radii also appear to induce star formation. The alignment of induced star-formation regions is observed in the hosts of some young radio sources (\citealt{Axon2000a, Labiano2008a}; see also contribution by O'Dea in this Volume). Induced star-formation near the jet stream is also seen in filaments along the jet of Centaurus A \citep{Santoro2015a}.

The difference between the passive star-formation rate, that is, the star-formation rate of the system before the introduction of a jet, and the jet-induced star-formation rate is more clearly seen in Figure~\ref{f:sfrd}, a map showing the ratio between the star-formation surface density at the first time step, before jet feedback, and at the last time step, after jet feedback.\footnote{This comparison can be made for these simulations because the gas consumption timescale and the cloud orbital timescale are much larger than the jet crossing time.} Regions in the inhomogeneous medium experiencing positive feedback are colored red, while regions experiencing negative feedback are in blue. Again, we see a strong alignment of the clumps of positive feedback in a cylindrical region about the main jet stream (or the jet axis), but we also note enhanced star-formation due to compression by the pressurized cocoon in the outer clouds that are still mostly intact. The net effect by the end of the simulation with a jet was an overall increase in SFR, but only by a modest 50\%{} compared to the passively star-forming case without a jet.

Whether positive or negative feedback dominates depends, in these simulations, on the size (or column density) of the clouds. We find that for cloud sizes greater than 25 pc, there is a phase in which positive feedback dominates, but if clouds are smaller than 25 pc, they are sufficiently quickly ablated for negative feedback to dominate. This is only an approximate division since the net effect also depends on the spatial distribution of the clouds. Furthermore, this analysis does not take into account gravitational instability as a condition for star-formation, and, numerically we cannot be sure that hydrodynamic ablation is properly captured. 

Galaxies in which the conditions for positive feedback by AGN jets may be optimal are high-redshift gas-rich disc galaxies \citep{Gaibler2014a}. If an AGN jet propagates perpendicular to the gas-rich galactic disc, its interaction depth is short (of order the disc height) resulting in comparatively weak negative feedback \citep{Sutherland2007a}. The high thermal pressure and ram pressure in the cocoons and bubble inflated by the jet can, however, compress the entire disc and cause galaxy-wide enhanced star-formation. \citet{Kim2012a} found an enhanced growth rate of the gravitational instability for pressure confined discs and an associated modified Toomre instability criterion. Combining this result with the theory of pressure-regulated star-formation \citep{Silk2009a}, \citet{Silk2013a} derived a modified SFR for a galactic disc confined by an AGN-driven pressure bubble that is proportional to the square root of the confinement pressure.

\citet{Gaibler2012a} performed three-dimensional hydrodynamic simulations of jets propagating perpendicular to a thick disc. After a brief phase in which the jet was confined while traversing the height of the disc, the jets emerged at different times into the halo, showing that the interaction of jets with a dense ISM can lead to asymmetries in the jet morphology. A ring of compressed gas and associated enhanced star-formation propagated cylindrically outward from the jet axis into the disc. The jets inflated cocoons and a bubble which, eventually, engulfed and pressurised the entire disc leading to enhanced galaxy-wide star-formation. By the end of the simulation, the SFR had risen by a factor of three since jet injection and was still rising. The enhanced SFR was in line with the predictions by \citet{Silk2013a}.

In neither the simulations by \citet{Gaibler2012a} nor ours, had the SFR converged by the end of the simulations. Clearly, positive feedback continues long after the jet has broken out of the dense regions of the ISM. Cloud compression at later times is driven partly by the ram-pressure of the turbulent back-flow of the jet and partly by the remaining overpressure in the bubble. Simulations on larger spatial scales and over longer timescales are required to determine the long-term total effect of positive versus negative feedback.

Since it is time-consuming to run three-dimensional jet simulations to study the efficiency of positive feedback over, say, at least a dynamical time of the galaxy (the relativistic jet requires time-steps to be very small), it is instructive to investigate the effect of over-pressurisation of a disc on its own. To this end, \citet{Bieri2015a} performed three-dimensional hydrodynamic simulations of self-gravitating gas-rich discs embedded in a hot, over-pressured halo and followed the evolution of the disc up to $\sim400$ Myr. Under the excess ambient pressure, the disc fragmented faster, leading to a greater number of clumps compared to control runs in which the disc was in pressure equilibrium with the disc. This was in agreement with the predictions by \citet{Kim2012a}. The SFR throughout the simulation was also enhanced as predicted by \citet{Silk2013a}.

\section{Summary}\label{s:summary}

We have reviewed selected results from hydrodynamical simulations of AGN jet feedback relevant to GPS and CSS radio galaxies. If dense gas exists in GPS and CSS radio galaxies, the relativistic jets are likely to interact strongly with the ISM, leading to jet frustration, multiphase outflows, and induced star-formation.

The difference in the outcomes of AGN feedback depends on whether the ISM clouds are mostly confined in a galactic disc or distributed more isotropically throughout the hot halo of the galaxy. In the former case, the thermal and ram pressure from the jet cocoon and jet-blown energy-bubble led to increased clump formation \citep{Kim2012a, Bieri2015a} and enhanced SFR \citep{Gaibler2012a, Silk2013a} in the disc. In the latter case, the jet is deflected and floods through the inter-cloud channels, carrying substantial ram pressure, ablating, dispersing, and accelerating material from the surface of the clouds, and clouds in bulk. For a given total mass and volume filling factor of dense clouds, negative feedback of this sort is more efficient the smaller the column densities of clouds are. Even in the case of isotropically distributed clouds, pressure-triggered star-formation may be important if clouds are sufficiently large so as not to be completely destroyed through ablation before having a chance to collapse gravitationally. Similar conclusions for strongly negative feedback in spherical cloud distributions and positive feedback in disc galaxies were found for energy-driven AGN disc winds \citep{Wagner2013a}.

Apart from estimating the SFR directly, the evolution of the density PDF of the dense phase is a simple way to assess the amount of cloud ablation and cloud coagulation, which approximate the amount of negative and positive feedback occurring during AGN feedback. 

The relevance of AGN jet feedback in the context of cosmological galaxy formation is still unclear. With ever-increasing data of radio galaxies from, e.g., LOFAR and ASKAP observations, and unprecedented detailed views of the ISM properties from ALMA observations, we are in need of more rigorous theoretical work involving realistic hydrodynamical simulations of AGN feedback in order to properly quantify the role of AGN jet feedback in the evolution of all galaxies that have undergone a radio-loud AGN phase.

\acknowledgements
The computations for this paper were undertaken on the NCI National Facility at the Australian National University. The software used in this work was in part developed by the DOE-supported ASC / Alliance Center for Astrophysical Thermonuclear Flashes at the University of Chicago. During the course of this research AYW was in part supported by a Japanese Society for the Promotion of Science (JSPS) fellowship (PE 11025). This work was supported in part by the {\it FIRST} project based on Grants-in-Aid for Specially Promoted Research by MEXT (16002003) and JSPS Grant-in-Aid for Scientific Research (S) (20224002). 

\bibliographystyle{an}


\end{document}